# Weighted SAMGSR: combining significance analysis of microarray-gene set reduction algorithm with pathway topology-based weights to select relevant genes

Suyan Tian[1,2*], Howard H. Chang[3], Chi Wang[4]

[1] Division of Clinical Epidemiology, The First Hospital of Jilin University, 71Xinmin Street, Changchun, Jilin, China, 130021
[2] School of Mathematics, Jilin University, 2699 Qianjin Street, Changchun, Jilin, China, 130012
[3] Department of Biostatistics and Bioinformatics, Rollins School of Public Health, Emory University, 1518 Clifton Road NE, Atlanta, GA, 30322
[4] Department of Biostatistics, Markey Cancer Center, The University of Kentucky, 800 Rose St., Lexington, KY, 40536

**\* Corresponding authors**

Email: windytian@hotmail.com

# Abstract

## Introduction

It has been demonstrated that a pathway-based feature selection method which incorporates biological information within pathways into the process of feature selection usually outperform a gene-based feature selection algorithm in terms of predictive accuracy, stability, and biological interpretation. Significance analysis of microarray-gene set reduction algorithm (SAMGSR), an extension to a gene set analysis method with further reduction of the selected pathways to their respective core subsets, can be regarded as a pathway-based feature selection method.

## Results and Discussion

In SAMGSR, whether a gene is selected is mainly determined by its expression difference between the phenotypes, and partially by the number of pathways to which this gene belongs, but ignoring the topology information among pathways. In this study, we propose a weighted version of the SAMGSR algorithm by constructing weights based on the connectivity among genes and then incorporating these weights in the test statistic.

## Conclusions

Using both simulated and real-world data, we evaluate the performance of the proposed SAMGSR extension and demonstrate that gene connectivity is indeed informative for feature selection.

# Introduction

Many researches have demonstrated that pathway-based feature selection algorithms, which utilize biological information contained in pathways to guide which features/genes should be selected, is usually superior to traditional gene-based feature selection algorithms in terms of predictive accuracy, stability, and biological interpretation [1–10]. Consequently, pathway-based feature selection algorithms have become increasingly popular and widespread.

Methods to incorporate pathway knowledge in feature selection algorithms can be classified into three categories – penalty, stepwise forward, and weighting. Their definitions and characteristics are presented in Figure 1. In the penalty category, an additional penalty term accounting for the pathway structure is added to the objective function for optimization. In essence, this penalty term provides some smoothness on nearby genes within a pathway, relying on the assumption that neighboring genes inside a pathway are more likely to function together or to be involved in the same biological process than those are far away. As a result, the 'driving' genes with subtle changes themselves but altering some of their neighbors' expression values dramatically are more likely to be selected. However, the theoretical complexity and computational intensity associated with penalization methods require special attention.

The second category of methods involve starting from one gene (e.g., the most significantly differentially expressed), and then adding genes gradually and evaluating on some statistic until no further gain upon this statistic can be obtained. For instance, the SAMGSR algorithm proposed by [11] belongs to this category, and it consists of two steps. Its first step is essentially an extension of SAM method [12] to all genes inside a pathway and the significance level of a pathway is determined using permutation tests. Then a core subset is extracted from each significant pathway identified by the first step using the magnitudes of individual genes' SAM statistics. Given that SAMGSR is based on two well-established techniques, i.e., SAM and permutation test, its real-world applications are expected to be numerous. Nevertheless, SAMGSR implicitly accounts for pathway knowledge by increasing the chance of those genes involved in many pathways being selected, which ignores pathway topology information. Most importantly, SAMGSR may miss those 'driving' genes with subtle change because the crucial criterion for a gene's inclusion in the reduced subsets is its gene expression difference between two phenotypes.

The third category is to create a pathway knowledge-based weight for each gene. For instance, the reweighted recursive feature elimination (RRFE) algorithm proposed by Johannes et al [13] used GeneRank [14] to calculate a rank for each gene and then weighed the coefficients before a support vector machine (SVM) model by this rank. It had been demonstrated that the resulting gene signatures have better stability and more meaningful biological interpretation[13]. Compared to the other two categories, while the weighting category is more straightforward, it has been underutilized. Its underutilization might be due to the estimated weights are subject to errors and biases where the impact on the resulting significant may be substantial.

In this study, we propose a hybrid method that combines SAMGSR with a pathway topology-based weight to carry out feature selection. As a mixture of the weighting method and stepwise forward method, the objective is to address some disadvantages associated with SAMGSR and the weighting method while borrowing their strengths. The proposed method is referred to as weighted-SAMGSR herein. Applying it to both the simulated data and real-world application, we evaluate if the weight reflecting gene connectivity information is valuable for feature selection.

# Materials and Methods

## Experimental Data

We considered two sets of microarray datasets in this study. One is for multiple sclerosis (MS) application and the other is for non-small cell lung cancer (NSCLC) application.

### MS Data

The MS application consisted of two microarray experiments. The first one included chips from the experiment E-MTAB-69 stored in the ArrayExpress [15] repository (http://www.ebi.ac.uk/arrayexpress). All chips were hybridized on Affymetrix HGU133 Plus 2.0 chips. In this study, there were 26 patients with relapsing-remitting multiple sclerosis (RRMS) and 18 controls with neurological disorders of a non-inflammatory nature. The second dataset was provided by the sbv IMPROVER challenge in the year of 2012 [16], which is accessible to the participants on the project website (http://www.sbvimprover.com). It was hybridized on Affymetrix HGU133 Plus 2.0, and there were 28 patients with RRMS and 32 normal controls.

### NSCLC Data

Microarray data used in the study are those deposited under accession numbers of GSE10245, GSE18842, GSE2109 (these three as the training set), and GSE43580 (the test set) in the Gene Expression Omnibus (GEO) repository. All experiments were hybridized on Affymetrix HGU133Plus 2.0 chips.

### Gene Sets

Gene sets were downloaded from the **Molecular Signatures Database** (MSigDB) [17]. In this study, we only considered c5 category. This category includes gene sets annotated by Gene Ontology (GO) terms. The current version (version 4.0) of MSigDB c5 category included 1,454 gene sets.

## Pre-processing Procedures

Raw data of E-MTAB-69 were downloaded from the ArrayExpress repository, and expression values were obtained using the fRMA algorithm [18] and normalization across samples was carried out using quantile normalization. The resulting expression values were on the $\log_2$ scale. When there were multiple probe sets representing the same gene, the one with the largest fold change was chosen. Raw data of the second set were downloaded from the sbv challenge website, and were separately pre-processed in the same way.

Raw data (CEL files) of all NSCLC data sets were downloaded from the GEO repository, and expression values were obtained using the fRMA algorithm [18]. Since the training set included data from different microarray experiments, the COMBAT algorithm (http://www.bu.edu/jlab/wp-assets/ComBat/Abstract.html) was used to eliminate batch effects. Again, the training set and the test set were pre-processed separately.

## Statistical Methods

### SAMGSR

SAMGSR extends a pathway analysis method called significance analysis of microarray-gene set (SAMGS) [19] to identify the 'core' subset for each significant gene set. In SAMGS, the following functional score is defined,

$$SAMGS_j = \sum_{i=1}^{|j|} d_i^2 \quad, \quad d_i = (\bar{x}_d(i) - \bar{x}_c(i)) / (s(i) + s_0)$$

where $d_i$ is the SAM statistic [12] and calculated for each gene involving in gene set $j$, $\bar{x}_d(i)$ and $\bar{x}_c(i)$ are the sample averages of gene $i$ for the diseased and control group, respectively. Parameter $s(i)$ is a pooled standard deviation that estimated by pooling all samples together, while $s_0$ is a small positive constant used to offset the small variability in microarray expression measurements, and $|j|$ represents the number of genes within gene set $j$. A gene set's significance is estimated using a permutation test by perturbing phenotype-labels.

For each significant gene set identified by SAMGS, the additional reduction step of SAMGSR partitions the entire set S into two subsets: the reduced subset $R_k$ including the first k genes and the residual one $\bar{R}_k$ for k=1,…, |j| by ordering the genes inside the set S decreasingly based on the magnitude of $d_i$. Then the significance level of $\bar{R}_k$ was evaluated. That is, let $c_k$ be the SAM-GS p-value of $\bar{R}_k$, the iteration stops when $c_k$ is larger than a pre-determined threshold for the first time. In our applications, we consider $c_k$ as a tuning parameter, and its optimal value is determined using a 5-fold cross-validation. Figure 2 provides a graph illustration on the SAMGSR algorithm.

## Statistical Metrics

As in previous study [20], we use four metrics - Belief Confusion Metric (BCM), Area Under the Precision-Recall Curve (AUPR), Generalized Brier Score (GBS), and misclassified error rate to evaluate the performance of a resulting gene signature. Our previous study and the references therein described those metrics in details. Briefly, they all range from 0 to 1. For BCM and AUPR the closer to 1 the better a classifier is, whereas the direction is opposite for GBS and misclassified error rate.

Besides these predictive performance statistics, we additionally include the Rand index to evaluate the stability or robustness of the resulting signatures. With k runs of an algorithm, Rand index is defined as

$$Rand = \frac{2}{k(k-1)} \sum_{i=1}^{k-1} \sum_{j=i+1}^{k} \frac{\bigcap(gs_i, gs_j)}{\bigcup(gs_i, gs_j)}$$

where $\bigcap$ represents the size of intersection between two gene lists and $\bigcup$ represents the size of union between two gene lists $gs_i$ and $gs_j$ obtained from the $i^{th}$ and $j^{th}$ runs. Certainly, Rand index can be defined at the level of pathways by replacing the gene lists with the pathway lists.

## Statistical Language and Packages

All statistical analysis was carried out in the R language version 3.1 (www.r-project.org).

# Results and Discussion

## Weight Construction and Weighted SAMGS Statistic

As described in the Introduction section, SAMGSR assumes that the number of gene sets within which a specific gene is contained is highly correlated with the pathway connectivity, namely, the more number of such gene sets is, the more number of other genes is connected to this gene. From the scatterplot (Figure 3), we found the correlation between the number of gene sets and the connectivity among genes is only moderate, indicating that SAMGSR does not fully account for connectivity information.

To tackle this drawback of SAMGSR, we propose to combine a weight constructed on the basis of connectivity information with the SAMGS statistic. Specifically for G genes under consideration, a G×G adjacency matrix is defined. Its $ij$ component $a_{ij}$ equals to 1 if genes $i$ and $j$ are connected, 0 otherwise. Because here we only consider an undirected pathway connectivity diagram, this adjacency matrix is symmetric. Then the connectivity weight for gene $i$ is defined as,

$$w_i = \sum_{j=1,\ldots,G} a_{ij}, \quad a_{ii} = 1$$

the connectivity information was retrieved from the protein-to-protein interaction (PPI) information provided by Human Protein Reference Database (HPRD). The PPI information was downloaded from the HPRD webpage (www.hprd.org), and then the adjacency matrix among genes was calculated using the R software.

In our proposed procedure, we include each gene's weight in its SAM statistic to obtain so-called weighted SAM and weighted SAMGS statistics and then replace SAM/SAMGS with their weighted counterparts to carry out pathway selection followed by individual gene selection. The proposed method is referred to as weighted-SAMGSR herein. In Fig 2, the definition of weighted SAMGS statistics and where they replace SAMGS statistics are presented. Within each specific pathway, SAMGSR ranks genes based on their SAM statistics and discards gene connectivity information completely. In contrast, the weighted-SAMGS algorithm weighs the genes with high connectivity more importantly. This is motivated to better detect 'driving' genes that are highly connected to other genes but have subtle expression differences to be selected increases by this means. In following subsections, we apply the weighted-SAMGSR algorithm to both simulated and real-world data.

In both SAMGSR and weighted-SAMGSR algorithms, $c_k$ is regarded as a tuning parameter that determines the sparseness of the final models. Its optimal values are determined via 10-fold CV by randomly dividing the whole training dataset into 10 roughly equal-sized folds. We apply either SAMGSR or weighted SAMGSR to 9 of these folds and verify their performance on the held-out fold. This step is repeated for each of the 10 folds as the held-out fold, and then the error rate is calculated. We then take the optimal $c_k$ value and apply SAMGSR or weighted SAMGSR to the whole training dataset to select genes in the final models, whose performances are evaluated using the independent test set. Of note, since for the SAMGSR methods the classifiers are not automatically produced along with the process of feature selection, we fitted support vector machine (SVM) models to estimate the corresponding coefficients of the selected genes.

## Simulated Data

Here, two simulations were used to characterize the weighted-SAMGSR algorithm and make comparisons with the SAMGSR algorithm. Here, we randomly chose 5 gene sets in the MSigDB c5 category. There are approximately 1000 genes inside these 5 gene sets. In the first simulation, we simulated the gene expression profile as independent random variables with a standard normal distribution and the sample size was 60. Then we simulated another set of normally distributed random variables and used it as the test set. In the second simulation, the observed expression values of the integrated NSCLC data were used to train the final model. The expression values were further normalized to have means of zeros and standard deviations of ones. The expression values of GSE43580 were used to test the final model and evaluate its performance.

We chose two genes—*HDAC1* and *GNAS* as the relevant genes and simulated the case and control groups using the following logit function,

$$\log it_{2vs1} = 0.37 X_{HDAC1} - 0.86 X_{GNAS}$$

because *HDAC1* has the higher connectivity, its coefficient was set as being smaller than that of *GNAS*. The simulation results are presented in Table 1.

Overall, the weighted-SAMGSR algorithm outperforms the SAMGSR algorithm. Namely, the weighted-SAMGSR algorithm has a substantially higher probability to identify HDAC1 whose signal is about 1.5 times weaker than GNAS and has better performance statistics in these two scenarios. Therefore, constructing the weights based on genes' topology information and incorporating those weights with the SAMGS statistics improve upon the feature selection capacity of SAMGSR.

## Real World Application

Previously, we applied SAMGSR to one set of multiple sclerosis (MS) data and showed that it possesses the feature selection trait, indeed [21]. In this study, we use the MS dataset again to evaluate if weighted-SAMGSR is superior to SAMGSR after accounting for the additional connectivity information among genes.

## MS Data

MS is the most prevalent demyelinating disease and the primary cause of neurological disability in young adults [22]. Here, we analyzed a set of MS real-world data to explore the discriminative capacity of expression profiles to separate MS patients from controls, and to characterize the proposed weighted SAMGSR method.

The results are presented in Tables 2 and 3. In Table 2, we observe the selected pathways by SAMGSR and weighted SAMGSR with high frequencies differ completely. On the level of individual genes, there are 6 overlapped genes. According to the genecards (www.genecards.org) database, two genes – POLD1 and MRE11A among these 6 genes are directly related with MS. In this application, weighted SAMGSR outperforms SAMGSR in terms of predictive performance, stability, and the number of directly related genes.

## NSCLC Data (>2 groups comparison)

Both the SAMGSR algorithm and the weighted SAMGSR algorithm can be adopted directly to deal with the multiple classes (>2 groups). Here, we used a set of NSCLC data to showcase this. In this application, the patients were categorized into four classes according to their respective histology subtypes and stages, i.e., adenocarcinoma at stage I (AC-I), adenocarcinoma at stage II (AC-II), squamous cell carcinoma at stage I (SCC-I), and squamous cell carcinoma at stage II (SCC-II). To classify on these four groups, we applied both SAMGSR algorithms twice—one for the subtype segmentation and one for the stage segmentation. Then the final posterior probabilities are P(AC-I)= P(AC)×P(stage I), P(AC-II)= P(AC)×P(stage II), P(SCC-I)= P(SCC)×P(stage I), and P(SCC-II)= P(SCC)×P(stage II), respectively. The results are given in Table 4.

Similar to the results from the sbv Lung cancer (LC) challenge [23], the segmentation between stages is not achievable whereas the segmentation between subtypes is good. Nevertheless, both SAMGSR algorithms identify more than 300 genes for the subtype segmentation while other studies had obtained similar performance using just one [24] or several genes[20]. We contribute this to the fact that SAMGSR is a filter method [25] that ignores the feature dependencies and thus tends to introduce all highly correlated features to the true relevant ones into the final model.

By accounting for the connectivity among genes, weighted SAMGSR outperforms SAMGSR in both applications, which is consistent with the results from the simulated data. Nevertheless, it is observed that such superiority differs in the two applications—being substantial in the MS application whereas marginal in the NSCLC application. This may be attributable to that many cancers are under intensive investigation and may be better curated in the major pathway databases. Therefore, the genes inside one specific pathway might be more likely to function together given the performance of SAMGSR in the NSCLC application is already good. In contrast, in the MS application the performance of SAMGSR is substantially inferior to that of those top 3 teams in the sbv challenge. In this case the extra information of gene connectivity becomes more valuable and makes the weighted SAMGSR algorithm comparable to the third team in this sub-challenge.

# Conclusions

Although SAMGSR is a pathway-based feature selection algorithm in nature, it treats all genes inside one pathway equally and assumes those 'core' genes in the pathway co-function together to regulate biological processes. To tackle its major drawback of discarding the gene topology knowledge completely, we propose a weighted extension to SAMGSR by creating the weights based on the connectivity among genes and combining those weights with the SAMGS statistics. Using simulations and real-world applications, we demonstrated that this weighted version of SAMGSR outperforms the original SAMGSR algorithm. In addition, the weight construction is very straightforward and has added no extra computing burden to the algorithm. Therefore, the weighted SAMGSR algorithm is preferred over the SAMGSR algorithm.

Pathway-based feature selection algorithms have become a topic of increasing interest in the field bioinformatics currently. Indeed, incorporating additional meaningful information does facilitate feature selection [26]. Nevertheless, the pathway knowledge is far from completeness and thus is subject to changes and errors, which limits the use of those pathway-based methods in practice. To address this, we suggest to analyze the real-world data with both gene-based methods and pathway-based methods and to explore if the pathway knowledge is informative for the specific application.

# Tables and Figures

## Table 1. Simulation results

|  | Training set | | Test set | | | |
|---|---|---|---|---|---|---|
| Method (Size[1]) | HDAC1(%) | GNAS(%) | Error (%) | GBS | BMC | AUPR |
| A. Simulated from 60 independent normal-distributed random variables | | | | | | |
| SAMGSR (3.8) | 19 | 100 | 16.5 | 0.118 | 0.733 | 0.921 |
| W-SAMGSR (6.23) | 65 | 100 | 13.2 | 0.101 | 0.755 | 0.948 |
| B. Simulated based on the NSCLC data | | | | | | |
| SAMGSR (3.94) | 0 | 100 | 44.5 | 0.256 | 0.517 | 0.550 |
| W-SAMGSR (6.28) | 77 | 100 | 40.5 | 0.241 | 0.534 | 0.621 |

Note: W-SAMGSR stands for weighted-SAMGSR; [1] stands for average the number of genes selected by either SAMGSR or W-SAMGSR over 100 replicates.

## Table 2. The selected pathways and genes on MS data

|  | Pathways with high frequency (frequency %) | Genes (frequency %) |
|---|---|---|
| SAMGSR | DNA Directed DNA Polymease Activity (100%)<br>DNA Polymease Activity (90%)<br>COVALENT_CHROMATIN_MODIFICATION (70%)<br>HISTONE_MODIFICATION (70%)<br><br>Stability=14.04% | **POLD4**(100%)  **POLD1**(80%)<br>PHB(80%)  **GPAA1**(70%)<br>**PIGT**(70%)  DPM3(70%)<br>**MRE11A**(60%)  **PI4KB**(60%)<br><br>Stability=12.83% |
| Weighted SAMGSR | DNA_RECOMBINATION(70%)<br>LIPOPROTEIN_BIOSYNTHETIC_PROCESS (70%)<br>NEGATIVE_REGULATION_OF_IMMUNE_SYSTEM_PROCESS(70%)<br>PROTEIN_AMINO_ACID_LIPIDATION (70%)<br>DEPHOSPHORYLATION(60%)<br>INOSITOL_OR_PHOSPHATIDYLINOSITOL_KINASE_ACTIVITY (60%)<br>LIPOPROTEIN_METABOLIC_PROCESS(60%)<br>PROTEIN_C_TERMINUS_BINDING (60%)<br><br>Stability=15.76 % | **MRE11A**(90%)  PTPRC(80%)<br>BRCA1(70%)  ATM(70%)<br>CHAF1A(70%)  **PIGT**(70%)<br>**GPAA1**(70%)  **PI4KB**(70%)<br>PEX16(60%)  **POLD1**(60%)<br>**POLD4**(60%)  PPP1CA(60%)<br><br>Stability=14.03% |

Note: Gene symbols in bold are those overlapped genes by SAMGSR and weighted SAMGSR; gene symbols

## Table 3. Performance statistics of selected genes on MS data

|  | Training set (10-fold CV results) | | | | Test set | | | |
|---|---|---|---|---|---|---|---|---|
| A. Performance of SAMGSR and weighted SAMGSR | | | | | | | | |
| Method (n) | Error (%) | GBS | BCM | AUPR | Error (%) | GBS | BCM | AUPR |
| SAMGSR (52) | 34.09 | 0.244 | 0.570 | 0.645 | 46.67 | 0.465 | 0.501 | 0.725 |
| W-SAMGSR (25) | 31.82 | 0.191 | 0.611 | 0.771 | 43.33 | 0.341 | 0.564 | 0.860 |
| B. Comparison with other feature selection algorithms | | | | | | | | |
| LASSO (30) | 34.09 | 0.275 | 0.632 | 0.672 | 46.67 | 0.377 | 0.499 | 0.747 |
| Penalized SVM(11) | 47.73 | 0.406 | 0.534 | 0.630 | 45 | 0.569 | 0.431 | 0.555 |
| C. Performance for the top 3 teams in sbv MS sub-challenge (among 54 teams) | | | | | | | | |
| Study (size) | Training data used/Method used | | | | Error (%) | GBS | BCM | AUPR |
| Lauria's (n>100) | E-MTAB-69/ Mann-Whitney test, then use top α % of the selected genes and Cytoscape to get the clusters on the test set | | | | -- | -- | 0.884 | 0.874 |
| Tarca's (n=2) | GSE21942 (on Human Gene 1.0 ST)/LDA | | | | -- | -- | 0.629 | 0.819 |
| Zhao's (n=58) | 7 other data and E-MTAB-69/Elastic net | | | | 30 | -- | 0.576 | 0.820 |

Note: W-SAMGSR: weighted SAMGSR; LDA: linear discrimination analysis. --: not available. Lauria's Tarca's and Zhao's studies [27–29] are the 3 best studies in the sbv MS sub-challenge.

**Table 4. Performance statistics of selected genes on NSCLC test data**

| | Training set (5-fold CV results) | | | | Test set | | | |
|---|---|---|---|---|---|---|---|---|
| A. Performance of SAMGSR and weighted SAMGSR | | | | | | | | |
| Method (n) | Error (%) | GBS | BCM | AUPR | Error (%) | GBS | BCM | AUPR |
| SAMGSR (30) [1] | 40.7 | 0.279 | 0.377 | 0.462 | 51.3 | 0.348 | 0.407 | 0.486 |
| W-SAMGSR (27)[1] | 37.2 | 0.276 | 0.378 | 0.453 | 51.3 | 0.345 | 0.405 | 0.492 |
| B. Comparison with other feature selection algorithms | | | | | | | | |
| LASSO (95) | 38.6 | 0.281 | 0.458 | 0.483 | 52.7 | 0.395 | 0.456 | 0.485 |
| pSVM (>100) | 42.8 | 0.370 | 0.344 | 0.428 | 53.3 | 0.433 | 0.385 | 0.397 |
| C. Performance for the top 3 teams in sbv NSCLC sub-challenge (among 54 teams) | | | | | | | | |
| Study (size) | Training data used/Method used | | | | Error (%) | GBS | BCM | AUPR |
| Ben-Hamo's (23) | GSE10245, GSE18842, GSE31799/PAM | | | | 49.3 | -- | 0.48 | 0.46 |
| Tarca's (25) | GSE10245, GSE18842, GSE2109/ moderated t-tests+LDA | | | | -- | -- | 0.459 | 0.454 |
| Tian's (66) | GSE10245, GSE18842, GSE2109/TGDR in hierarchical way | | | | 53.3 | 0.374 | 0.440 | 0.471 |

Note: W-SAMGSR: weighted SAMGSR; pSVM: penalized support vector machine (SCAD penalty term); LDA: linear discriminant analysis; PAM: partitioning around medoid; TGDR: threshold gradient descent regularization; [1]The sizes of final model for the stage segmentation because the results for the subtype segmentation for both algorithms are identical (but the final size>300). Ben-Hamo's study [24],Tarca's study [28]and Tian's study [30] are the 3 best studies in the sbv LC sub-challenge.

**Figure 1. Three categories of pathway-based feature selection algorithms.** The filter and embedded methods are two typical types for the gene-based feature selection algorithms. As defined by [25], filter methods access the relevance of features by calculating some functional score while embedded methods search for the optimal subset simultaneously with the classifier construction.

| Category/description | Property | Pathway topology information | Examples [Ref.] |
| --- | --- | --- | --- |
| Penalty: add an extra penalty term which accounts for the pathway structure to the objective function, then optimize the resulting function to get the final gene subset | Embedded feature selection methods, carry out feature selection and coefficient estimation simultaneously, moderate to heavy computing burden | Need the pathway topology information for all genes, e.g., are they connected and the distance between them | Net-Cox [Zhang et al, 2013] netSVM [Chen et al, 2011] |
| Stepwise forward: order genes based on one specific statistic, and then add gene one by one until there is no gain on the pre-defined score. | Usually filter methods, the beneath concepts and theory are simple. However, they also inherits the filter methods' drawbacks of inferior model parsimony and thus high false positive rate. | Usually ignore the pathway topology information, the decision hinges mainly on the genes' expression values | SAM-GSR [Dinu et al, 2009] SurvNet [Li et al, 2012] |
| Weighting: create some kind weights according to the pathway knowledge and then combine with other feature selection methods to identify the relevant genes | With different weights, the chance of those "driving" genes with subtle change being selected increases. However, if the estimated weights subject to big biases, the resulting model might even be inferior to those without weights. | Account for the pathway topology information. | RRFE [Johannes et al, 2010] DRW [Liu et al, 2013] |

# Figure 2. Diagrams to elucidate both SAMGSR and weighted SAMGSR algorithms.

A. Diagrams on SAMGSR and weighted SAMGSR algorithms

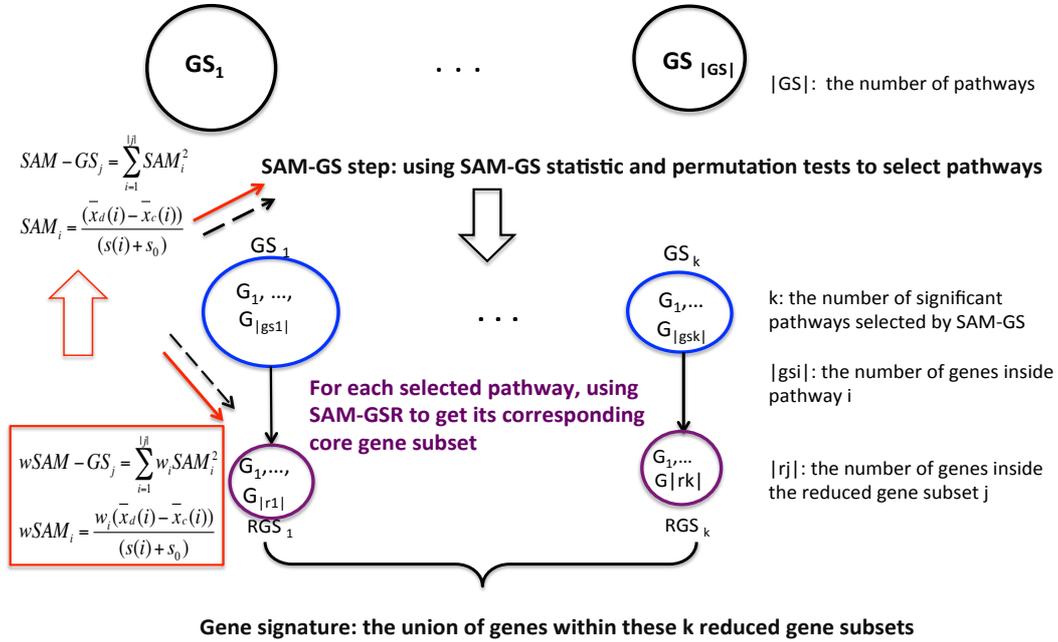

B. Definition of pathway topology-based weights

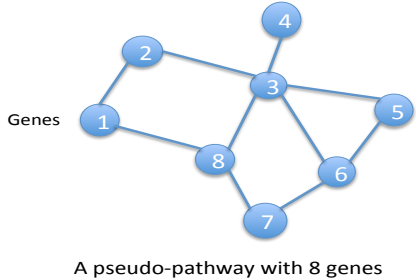

A pseudo-pathway with 8 genes

Adjacency matrix

|    | g1 | g2 | g3 | g4 | g5 | g6 | g7 | g8 | w |
|----|----|----|----|----|----|----|----|----|---|
| g1 | 1  | 1  | 0  | 0  | 0  | 0  | 0  | 1  | 3 |
| g2 | 1  | 1  | 1  | 0  | 0  | 0  | 0  | 0  | 3 |
| g3 | 0  | 1  | 1  | 1  | 1  | 1  | 0  | 1  | 6 |
| g4 | 0  | 0  | 1  | 1  | 0  | 0  | 0  | 0  | 2 |
| g5 | 0  | 0  | 1  | 0  | 1  | 1  | 0  | 0  | 3 |
| g6 | 0  | 0  | 1  | 0  | 1  | 0  | 1  | 0  | 3 |
| g7 | 0  | 0  | 0  | 0  | 0  | 1  | 1  | 1  | 3 |
| g8 | 1  | 0  | 1  | 0  | 0  | 0  | 1  | 1  | 4 |

$$w_i = \sum_{j=1,\ldots,G} a_{ij}$$

**Figure 3. Scatterplot of the number of gene sets involved versus the gene connectivity.** ρ is the estimated Spearman correlation coefficient between the number of gene sets involved and (1+the number of connected genes)

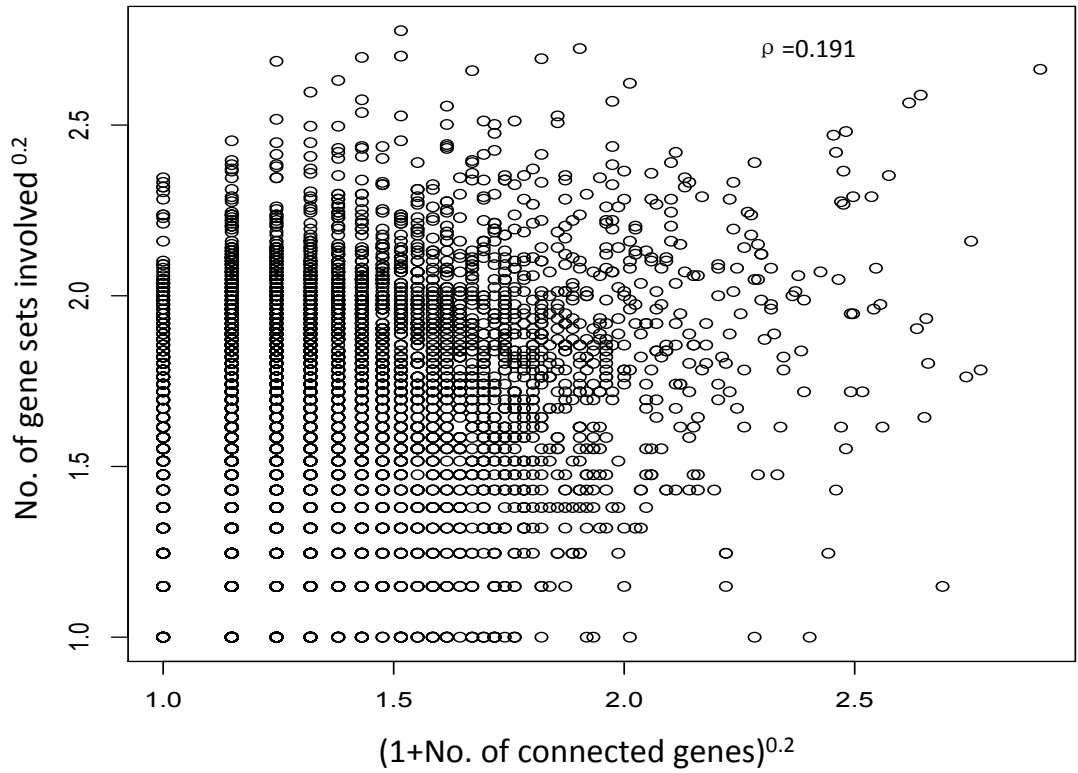